\documentstyle[aps,epsf]{revtex}
\newcommand{\be}{\begin{eqnarray}}
\newcommand{\ee}{\end{eqnarray}}
\newcommand\del{\partial}

\newcommand\Dels{D \hspace{-0.26cm} /}

\newcommand{\lton}{\mathrel{\lower.9ex
                  \hbox{$\stackrel{\displaystyle <}{\sim}$}}}

\begin{document}


\title{
\begin{flushright} 
{\small BNL-71005-2003-JA}\\
{\small SUNY-NTG-02/33}
\end{flushright} 
Melting the Diquark Condensate in Two-Color QCD:\\  
A Renormalization Group Analysis} 
\author{J. Wirstam$^{(a)}$\footnote{{\sl Former address: Department of
Physics, Brookhaven National Laboratory, Upton, NY 11973, USA}},  
J. T. Lenaghan$^{(b)}$, and K. Splittorff$^{\,(c)}$}

\address{$^{(a)}$ Swedish Defense Research Agency (FOI), S-172 90
 Stockholm, Sweden \\
 $^{(b)}$ Department of Physics, University of Virginia, 382 McCormick Rd., Charlottesville, {\sl VA 22904-4714} \\
 $^{(c)}$ Department of Physics and Astronomy, SUNY, Stony Brook, {\sl New York 11794}} 
\maketitle

\begin{abstract}
We use a Landau theory and the 
$\epsilon$-expansion to study the superfluid phase transition 
of two-color QCD at nonzero temperature, $T$, and 
baryonic chemical potential, $\mu$. At low
$T$, and for $N_{\! f}$ flavors of massless quarks, the global
SU(N$_{\!f}$)$\times$SU(N$_{\!f}$)$\times$U(1) symmetry is spontaneously
broken by a diquark condensate down to Sp(N$_{\!f}$)$\times$Sp(N$_{\!f}$)
for any $\mu > 0$.  As the temperature increases, the diquark condensate
melts, and at sufficiently large $T$ the symmetry is restored. Using
renormalization group arguments, we find that in the presence of 
the chiral anomaly term there can be a second order phase transition 
when N$_{\!f}=2$ or N$_{\!f}\geq 6$, while the transition is 
first order for 
N$_{\!f}=4$.  We discuss the relevance of these 
results for the emergence of a tricritical point recently observed 
in lattice simulations.
\end{abstract}

\section{Introduction}
\label{I}

During the last few years, a considerable amount of interest has been
devoted to the phases of QCD at a nonzero baryonic chemical potential
$\mu$. Based on either one-gluon exchange or instanton induced
interactions, there are compelling theoretical indications that a
direct attractive quark-quark channel gives rise to diquark
condensation at asymptotically large $\mu$, similar in spirit to the
BCS-scenario \cite{csc}. 
In particular, very dense quark matter is unstable to the formation of a 
diquark condensate, and thus is generally believed to be a 
color superconductor at large $\mu$.
The QCD-based predictions rely on perturbation theory
and are trustworthy when the coupling constant is small, 
$g(\mu)\ll 1$, i.e. $\mu \gg \Lambda_{{\rm QCD}}$, where
$\Lambda_{{\rm QCD}}$ is the hadronic scale (see \cite{RS}).  To
analytically study  
the properties of QCD at more moderate densities, where the quark
chemical potential is only a few times 
$\Lambda_{{\rm QCD}}$, one has to rely on other methods.  For instance, one can
resort to phenomenological models that are inspired by, but not
derived from, QCD. An alternative approach would be to use numerical
simulations, since lattice gauge theory is a first-principle method
capable of making quantitative predictions of strong-coupling QCD.  At
$T>0$ and $\mu =0$, the numerical results show that there is a phase
transition between the confined and deconfined phases of pure glue
matter, at a critical temperature $T_c \simeq 270$ MeV
\cite{lattice1}. With massless quarks, there is a transition to a
chirally symmetric phase at $T_c \simeq 155-175$ MeV \cite{lattice2},
and recent simulations also show that deconfinement occurs
simultaneously with the chiral phase transition \cite{lattice3}.
However, at a nonzero chemical potential the results have so far been
very limited, due to the severe problems lattice gauge theory
encounters at nonzero $\mu$.  These difficulties arise from the fact
that the action in the SU(3) gauge theory becomes complex-valued when
$\mu \neq 0$ \cite{kogut1}, which in turn inhibits the importance
sampling in the Monte-Carlo method.  Although different computational
schemes have been implemented
\cite{newschemes,Alford:1998sd,Chandrasekharan:1999cm,FodorKatz} 
the problem with large--scale simulations at nonzero chemical remains
unsettled. 

For $N_c=2$, the situation is much more favorable for numerical
simulations. In this case, the fundamental representation is
pseudoreal, and as a consequence the lattice measure remains
real-valued even at $\mu \neq 0$ \cite{realdet}. Since the fermion
determinant is real, it is possible to perform lattice calculations
for an even number of flavors using the standard Monte-Carlo
algorithms. Although two-color QCD is worthwhile to study in its own
right, the fact that it is a four-dimensional, confining non-Abelian
theory makes it particularly interesting as a toy model for real
QCD. Naturally, some caution is needed, since while there are some
important commonalities between two-- and three--color QCD, there are
some striking differences. For example, hadron spectroscopy will be
completely unrelated since for $N_c=2$ two quarks can form a color
singlet.  Likewise, the diquark condensate is a color 
singlet for $N_c=2$, while 
for $N_c=3$ it is not.
Finally, in contrast to the color superconducting state of 
dense quark matter for $N_c=3$, diquark condensation for $N_c=2$ 
only gives rise to superfluidity. 

Despite these differences, two-color QCD is still useful for
testing general ideas and comparing analytical calculations with
lattice simulations. Indeed, early lattice studies of two-color
QCD yielded interesting results at both nonzero chemical potential and
temperature \cite{realdet}. In recent years,
an extensive numerical program has been undertaken and a consistent 
picture of two--color QCD at nonzero $T$ and $\mu$ has emerged
\cite{Kogut:2001if,Kogut:2001na}.  
For $T=0$, as the chemical
potential is increased, a colorless diquark
condensate, $\langle \psi \psi \rangle$, forms at 
$\mu_c \simeq m_{\pi}/2$, where $m_{\pi}$ 
is the pion mass.  For 
$\mu \geq \mu_c$, the condensate spontaneously breaks baryon number 
and gives rise to
superfluidity. This phase transition is most likely second
order, at least for one and two flavors of staggered
quarks for which the simulations have been performed. 
Alternatively, two--color QCD can be studied 
analytically using chiral perturbation theory. 
This effective theory describes the dynamics of the
Goldstone modes associated with the spontaneous breaking 
of chiral symmetry and is valid for energies well below the chiral 
symmetry breaking scale.
In particular, the critical baryonic chemical potential, 
$\mu_c=m_\pi/2$, is well within the effective range and indeed the 
onset of baryon density and diquark condensation at zero temperature
is fully understood in this context at both
the classical \cite{RSSV,KST,KSTVZ,SS,SSS,KT} and one--loop level
\cite{STV1}. 
The analytical predictions agree well with the lattice results, 
thus further strengthening our current
understanding as well as showing the great utility of 
having both analytical and numerical tools available.

Although the different phases and associated phase transitions at $T=0$
and $\mu \neq 0$ have been investigated in some detail, less is
known about the phase structure of two--color QCD in the $(\mu,T)$-plane. 
Analytical results employing renormalization group arguments
for the chiral symmetry restoring phase transition at nonzero
temperature but vanishing chemical potential were presented in Ref.\
\cite{wirstam}, and the phase diagram at nonzero $T$ and $\mu$ has
been studied within a random matrix model \cite{rmm}. On the lattice,
simulations corresponding to $N_{\! f}=4$ continuum flavors indicate
that as $T$ increases at a fixed moderate ratio $x = 2\mu /m_\pi$, 
the system undergoes a second order phase transition where
$\langle \psi \psi \rangle$ vanishes \cite{Kogut:2001if}.  As this
ratio increases, this line of second order phase
transitions passes through a tricritical point and the phase
transition becomes first order \cite{Kogut:2001if,latticemut2}. 
Recently the one loop calculation within chiral perturbation theory was
extended to nonzero temperature \cite{STV1,STV2}, and the temperature
dependence of the critical chemical potential for $\mu \sim m_\pi/2$
was calculated.  Furthermore, the presence of a tricritical endpoint
of this line of second order phase transitions was confirmed. The
tricritical point was predicted to occur at $T\sim m_\pi$ \cite{STV2}
in qualitative agreement with the lattice results \cite{Kogut:2001if}.


In this paper we further study the phase diagram at nonzero
temperature and baryonic chemical potential by using a
Landau theory combined with one--loop
renormalization group arguments \cite{BKM}. The Landau theory 
is fundamentally different from the chiral perturbation theory analysis
in that it attempts to include all modes which become massless at the
critical temperature.  In addition to being a different approach, we
are also considering a different range of the $(\mu,T)$-plane than in
Ref.\ \cite{STV2}.  Here, we study the chiral limit with a nonzero
baryonic chemical potential. Since we consider the deep diquark phase
a large temperature limit is of interest (in the chiral perturbation
theory approach the low temperature limit was used for $\mu\sim
m_\pi/2$).  Our motivation is twofold. First of all, as mentioned above,
the phase diagram of two-color QCD needs to be better understood, and
one of our objectives is to provide a step in that direction.  In
particular, the emergence of the tricritical point at intermediate
values of $T$ and $\mu$ requires further explanation.  Although the
discussion will be concerned with the case of massless quarks,
$m_q=0$, for simplicity, our results and predictions are expected to
apply also for nonzero degenerate quark masses as long as $N_{\! f}$
is even and $\mu\gg m_\pi/2$. Secondly, further comparisons between
different analytical approaches and numerical simulations on the
lattice will hopefully increase our understanding of both methods, and
lay the groundwork for future attempts in full $N_c=3$ QCD at nonzero
baryon density.

We assume throughout that the gauge group is $SU(2)$ and that there
are an even number $N_{\! f}$ of massless flavors.  
In the next section, we discuss the pseudoreal
representation of $SU(2)$ and its implications for the $N_c=2$ QCD
action and the symmetry breaking patterns.  We then incorporate these
properties into an effective Lagrangian.  In Sec.\ III, we present our
renormalization group analysis and discuss the order of the phase
transition for various $N_{\! f}$.  We first assume the absence of the
chiral anomaly and later consider its implications.  We end with our
conclusions and an outlook in Sec. IV.

\section{Two--Color QCD and Effective Theories}
\label{II}

\subsection{The microscopic action}

The matter part of the QCD action with massless quarks and 
at zero chemical potential is 
\be
S_0= -i \sum_{f=1}^{N_{\! f}} \int d^4 \! x \, \bar{\psi}_f^i \left ( \del_{\mu} \delta^{j}_i -ig A_{\mu}^a 
(T^a)^{j}_i \right ) 
\gamma^{\mu} \psi_{jf} = -i \sum_{f=1}^{N_{\! f}} \int d^4 \! x \, \bar{\psi}_f \Dels \, \psi_f \ , \label{startaction}
\ee
where the Minkowski metric is $g_{\mu \nu} = {\rm diag} (+---)$ and the
color matrices are given by $T^a = \sigma^a /2$, with $\sigma^a$ the usual Pauli matrices.
Defining $ \sigma_{\mu} = (1, \vec{\sigma})$ and $\bar{\sigma}_{\mu} = (1, -\vec{\sigma})$, 
we have in the two-component formalism,     
\be
\psi_{if} \equiv \left ( \matrix{\xi_{\alpha} \cr \bar{\lambda}^{\dot{\alpha}} } \right )_{if} \ \ \ , \ \ \ \ 
\gamma^{\mu} \equiv \left ( \matrix{ 0 \, & \sigma^{\mu} \cr \bar{\sigma}^{\mu} \, & 0} \right ) \ .
\ee
Rewriting the action in Eq.\ (\ref{startaction}) in this Weyl notation gives, after a partial integration,
\be
S_0= -i \sum_{f=1}^{N_{\! f}} \int d^4 \! x \, \biggl [ \bar{\lambda}_{jf} \bar{\sigma}^{\mu} \left ( \del_{\mu} \delta^j_i 
+ig A_{\mu}^a (T^a)^j_i \right ) \lambda_f^i + \bar{\xi}^i_f \bar{\sigma}^{\mu} \left ( \del_{\mu} \delta^j_i 
-ig A_{\mu}^a (T^a)^j_i \right ) \xi_{jf} \biggr ] \ . \label{weylaction}
\ee
Using the pseudo--reality relation in SU(2), $\sigma_2 \sigma_a \sigma_2 = -\sigma^T_a$,
where the superscript $T$ denotes the transpose, the interaction term for $\lambda$ 
can be reorganized as
\be 
\bar{\lambda}_{jf} (T^a)^j_i \lambda_{f}^i = \bar{\lambda}_f T_a^T \lambda_f = 
-(\bar{\lambda}_f\sigma_2 ) T_a (\sigma_2 \lambda_f) \ .
\ee
Since $\lambda_f$ transforms as $\bar{{\bf 2}}$ under a gauge transformation,
$\sigma_2 \lambda_f$ transforms as $\xi$, i.e. like {\bf 2}.
Thus, by making the following flavor decomposition within the same color representation,
\begin{equation}
Q_f \equiv \left\{ \matrix{
\xi_f  & f=1,\ldots ,N_{\! f} \cr \sigma_2 \lambda_f   &  
        \,\,\,\,\,\,\,\,\,\,\,\ \ \  
        f=N_{\! f} +1,\ldots ,2N_{\! f} } \right . \ \ \ , \ \ \ 
\bar{Q}_f \equiv \left \{ \matrix{ \bar{\xi}_f  & f=1,\ldots ,N_{\! f} \cr
\bar{\lambda}_f \sigma_2 & \,\,\,\,\,\,\,\,\,\,\,\ \ \  
f=N_{\! f} +1,\ldots ,2N_{\! f}  } \right . \ ,
\end{equation}
we can write the action in Eq.\ (\ref{weylaction}) as,
\be
S_0=-i \sum_{f=1}^{2N_{\! f}} \int d^4 \! x \, \bar{Q}_f \left ( \del_{\mu} -igA_{\mu}^aT^a \right ) 
\bar{\sigma}^{\mu} Q_f \ .
\label{enlargedaction}
\ee
As can be seen from Eq.\ (\ref{enlargedaction}), the na{\"{\i}}ve
global symmetry 
SU($N_{\! f}$)$_L \times$SU($N_{\! f}$)$_R \times$U(1)$_B$ is in fact enlarged to SU(2$N_{\! f}$) in two-color QCD,
as a direct consequence of the pseudo-reality condition. 

However, this enlarged symmetry exists only as long as $\mu =0$. 
By introducing 
a common nonzero baryonic chemical potential $\mu$, 
the action in Eq.\ (\ref{startaction}) is augmented by an additional term,
\be
S_0 \stackrel{\mu > 0}{\longrightarrow}
S = S_0 - \mu \sum_{f=1}^{N_{\! f}} \int d^4 \! x \, \bar{\psi}_f^i \gamma^0 \psi_{if} = S_0- \mu \int d^4 \! x \,
\bar{Q}{\bf B}Q \ , \label{addmu}
\ee
where ${\bf B}$ is the (2$N_{\! f}$)$\times$(2$N_{\! f}$) diagonal 
baryonic charge matrix,
\be
{\bf B} \equiv \left ( \matrix{ {\bf 1} & {\bf 0} \cr {\bf 0} & -{\bf 1} } \right ) \ ,  \label{defineB}  
\ee
and ${\bf 0}$ and ${\bf 1}$ are $N_{\! f}\times N_{\! f}$ matrices.
The $\mu$-dependent term is not an SU(2$N_{\! f}$) singlet, since the block-diagonal form is
only invariant under rotations that do not mix the different blocks. In other words, 
when $\mu\neq 0$ the global symmetry is reduced to the more familiar
SU($N_{\! f}$)$_L \times$SU($N_{\! f}$)$_R \times$U(1)$_B$. 
Furthermore, in the presence of a common quark mass, $m_q$, the symmetry is reduced even further, since 
the mass term explicitly breaks the axial symmetry.  The global symmetry in this case is only
SU($N_{\! f}$)$_V \times$U(1)$_B$.


\subsection{Spontaneous breaking of symmetries} 

When $\mu=m_q=0$, we assume, as indicated by lattice simulations
\cite{Kogut:2001if,Kogut:2001na}, 
that the chiral SU(2$N_{\! f}$) symmetry
is spontaneously broken at low temperatures. A Lorentz and gauge invariant chiral condensate must be of 
the form \cite{peskin}
\be
\bar{\psi} \psi \propto Q_i {\bf E}^{ij}Q_j \ ,
\ee
with $i$ and $j$ the flavor indices and ${\bf E}$ a (2$N_{\! f}$)$\times$(2$N_{\! f}$) antisymmetric matrix.
Assuming a maximal flavor symmetry scenario \cite{peskin}, the standard form for ${\bf E}$ is given by
\be
{\bf E} \equiv \left ( \matrix{ {\bf 0} & {\bf 1} \cr - {\bf 1} & {\bf 0} } \right ) \ .
\ee
If the chiral condensate is nonvanishing, the global symmetry is spontaneously broken to the subgroup 
that leaves ${\bf E}$ invariant, namely Sp(2$N_{\! f}$),
and the symmetry breaking pattern SU(2$N_{\! f}$)$\rightarrow$Sp(2$N_{\! f}$) gives rise to $(2N_{\! f}^2-N_{\! f}-1)$ Goldstone bosons.

On the other hand, when $\mu >0$ and the quark mass $m_q\neq 0$, there
is only a SU($N_{\! f}$)$_V \times$U(1)$_B$ symmetry. Although the
chiral condensate does not break any of these symmetries, the lattice
results show that the preferred ground-state configuration is still
the chirally broken one in the region where $\mu \leq m_{\pi}/2$
\cite{Kogut:2001if,Kogut:2001na}.  
However, at a baryonic chemical potential 
$\mu = m_{\pi}/2$, on-shell baryons are produced out of the vacuum, the
chiral condensate begins to decrease and a diquark condensate appears:
\be
\psi \psi \propto Q^T {\bf \Sigma} Q \ ,
\ee
where ${\bf \Sigma}$ is a (2$N_{\! f}$)$\times$(2$N_{\! f}$) matrix given by
\be
{\bf \Sigma} \equiv i \left ( \matrix{ {\bf I} & {\bf 0} \cr {\bf 0} & {\bf I} } \right ) \ , \ \ \ \ \ \ {\rm with} \ \ \ \ \ \ 
{\bf I} \equiv \left ( \matrix{ {\bf 0} & -{\bf 1} \cr {\bf 1} & {\bf 0} } \right ) \ ,
\ee
and ${\bf I}$ is an $N_{\! f}\times N_{\! f}$ matrix. 
A nonvanishing diquark condensate ($m_q\neq0$ and $\mu\geq m_\pi/2$) breaks the 
SU($N_{\! f}$)$_V \times$U(1)$_B$ symmetry down to Sp($N_{\! f}$) and gives
rise to a Bose condensed phase.

When the quark mass $m_q=0$, $m_{\pi}$ also vanishes. This situation requires 
some additional considerations
for the diquark condensate.  
There is an interesting relationship, 
peculiar to two-color QCD, between the chiral and diquark condensates, at
$\mu =0$. Consider the particular flavor transformation 
matrix $M_1\!\in \,$SU(2$N_{\! f}$),
\be
M_1= \exp\left [ (3\pi i/4){\bf X}\right ] \ ; \ \ \ \ \ {\bf X} = 
\left ( \matrix{ {\bf 0} & {\bf I} \cr -{\bf I} & {\bf 0} } \right ) \ .
\ee
Under an SU(2$N_{\! f}$) rotation, the chiral condensate transforms as 
\be
{\bf E} \rightarrow 
M_1{\bf E}M_1^T = {\bf \Sigma} \ .
\ee
Hence, there is no distinction between the chiral and diquark condensate, and
due to this identity we may as well define the symmetry breaking at $\mu=0$ (and $m_q=0$) to be caused by
a diquark condensate. There is no difference, and
the breaking pattern remains SU(2$N_{\! f}$)$\rightarrow$Sp(2$N_{\! f}$).
(For further discussion see \cite{DP}.)


\subsection{The Landau theory at $\mu=0$}

To model the $m_q=0$ symmetry breaking, and its restoration as the temperature is increased 
for $\mu =0$, one can start from
a Landau theory \cite{wirstam}. 
We assume that the Landau field can be parameterized by 
an antisymmetric, complex-valued
(2$N_{\! f}$)$\times$(2$N_{\! f}$) matrix $\Phi$, $\Phi \sim QQ^T$, that 
transforms under the SU(2$N_{\! f}$) symmetry as $\Phi \rightarrow M\Phi M^T$ 
with $M\!\in \,$SU(2$N_{\! f}$). The effective Lagrangian\footnote{We thank F.\ Sannino for access to his notes in which this effective Lagrangian was first written down.} which is renormalizable and invariant under the global SU($2N_{\! f}$) flavor symmetry is
\be
L_0 = \frac{1}{2} {\rm Tr}\left [ (\del_{\mu}\Phi^{\dagger} ) (\del^{\mu}\Phi ) \right ]
 - \frac{m^2}{2} {\rm Tr} \left [\Phi^{\dagger} \Phi \right ] -\lambda_1 \bigl ( {\rm Tr} 
\left [ \Phi^{\dagger} \Phi \right ]
\bigr )^2 - \lambda_2 {\rm Tr} \bigl [ \left (\Phi^{\dagger} \Phi \right )^2 \bigr ] 
- c \left [ {\rm Pf} (\Phi) + {\rm Pf} (\Phi^{\dagger}) \right ] \ . \label{renormlag}
\ee
The last term in Eq.\ (\ref{renormlag}) is the so-called Pfaffian \cite{SV},
\be
{\rm Pf} (\Phi) = \frac{1}{2^{N_{\! f}} N_{\! f}!} \sum_P (-1)^P \Phi_{i_1 i_2} 
\cdots \Phi_{i_{(2N_{\! f}-1)} i_{(2N_{\! f})}}  \, \, ,
\ee
where $i_k = 1,\ldots ,2N_{\! f}$, and $P$ denotes a summation over all permutations of 
$\{ i_1,\ldots ,i_{2N_{\! f}} \}$ 
with $(-1)^P$ the sign of the permutation. This term ensures that the axial U(1)$_A$ is explicitly
broken, since if $c=0$ the Lagrangian is actually invariant under the larger U(2$N_{\! f}$) group.
Stability of the classical potential
requires that $\lambda_2 \geq 0$ and $\lambda_1 +\lambda_2/(2N_{\! f})\geq 0$, and by choosing
$m^2<0$ at $T=0$ the potential is minimized by a nonzero expectation
value $\Phi_0 = \rho_0 {\bf \Sigma}$, with 
$\rho_0^2 = -m^2/(8N_{\! f}\lambda_1+4\lambda_2)$ for $c=0$. This gives the correct symmetry breaking
pattern SU(2$N_{\! f}$)$\rightarrow$Sp(2$N_{\! f}$), where we for definiteness 
took $\Phi_0$ to be proportional to ${\bf \Sigma}$ instead of ${\bf E}$.
As discussed above, these are equivalent choices.

\subsection{The Landau theory at $\mu\neq0$}

We now introduce a nonzero baryonic chemical potential into the
effective theory.  The global symmetry SU(2$N_{\! f}$) is reduced to
SU($N_{\! f}$)$_L \times$SU($N_{\! f}$)$_R \times$U(1)$_B$ and
consequently there is a distinction between the chiral and the diquark
condensate.  Since we set $m_{\pi}=0$, the ground state has 
$\langle \psi \psi \rangle \neq 0$ and 
$\langle \bar{\psi} \psi \rangle =0$ at
any $\mu > 0$, breaking the symmetry spontaneously down to 
Sp($N_{\! f}$)$\times$Sp($N_{\! f}$).  These aspects must 
of course be reflected
in the effective Lagrangian when a chemical potential is added to
it. A general discussion of Landau theories and 
the $\epsilon$-expansion at nonzero chemical potential is given 
in Ref.\ \cite{SLW}. The important point to note here is that due to the
pseudoreality of the representation, 
the order parameter is the same\footnote{We consider 
values for the chemical potential which
are not asymptotically large so that the effective degrees 
of freedom are given by bosonic diquark bound states 
and not by the quarks and gluons themselves. A more restrictive
upper bound on $\mu$ is given by half of the mass of the lightest vector 
meson. At this value of the chemical potential, 
the formation of a vector condensate is expected
\cite{LSS,Sannino:2001fd,Alles:2002st}.}  as at $\mu =0$, namely $QQ^T$. 
As in mean field theory, we assume that the quartic coupling constants
remain unaffected by the introduction of $\mu$, although we will later
consider the possible influence of $\mu$ on the coupling $c$ in Eq.\ (\ref{renormlag}). 
The dependence of the Lagrangian
on $\mu$ is fixed uniquely by the symmetries and renormalizability.
The transformation
properties of the order parameter field is identical to that of the
Goldstone field in chiral perturbation theory. Hence, the coupling of
$\mu$ to $\Phi$ enters through the covariant derivative
\cite{KST}
\be
\del_{\nu}\Phi &&\rightarrow \del_{\nu}\Phi -i\mu \delta_{\nu 0}\left ( {\bf B}\Phi + \Phi {\bf B} \right ) \ ,
\nonumber \\ 
\del_{\nu}\Phi^{\dagger} &&\rightarrow \del_{\nu}\Phi^{\dagger} +i\mu \delta_{\nu 0}\left ( {\bf B}
\Phi^{\dagger} + \Phi^{\dagger} {\bf B} \right ) \ .
\ee
The Lagrangian in Eq.\ (\ref{renormlag}) then becomes
\be
L_0 \rightarrow L = &&\frac{1}{2} {\rm Tr}\left [ (\del_{\mu}\Phi^{\dagger} ) (\del^{\mu}\Phi ) \right ]
+i\frac\mu2 {\rm Tr}\left [ (\del_0\Phi )(\Phi^{\dagger}{\bf B}+{\bf B}\Phi^{\dagger})-h.c.\right ] +\frac{\mu^2}{2} {\rm Tr}
\left [ ({\bf B}\Phi + \Phi {\bf B})({\bf B}\Phi^{\dagger} + \Phi^{\dagger} {\bf B})\right ] 
 \nonumber \\ &&- \frac{m^2}{2} {\rm Tr} \left [\Phi^{\dagger} \Phi \right ]-\lambda_1 \bigl ( {\rm Tr} 
\left [ \Phi^{\dagger} \Phi \right ]
\bigr )^2 - \lambda_2 {\rm Tr} \bigl [ \left (\Phi^{\dagger} \Phi \right )^2 \bigr ] 
- c \left [ {\rm Pf} (\Phi) + {\rm Pf} (\Phi^{\dagger}) \right ] \ . \label{lagwithmu}
\ee
Given that $\Phi$ transforms as $\Phi\rightarrow M\Phi M^T$,
the Lagrangian in Eq.\ (\ref{lagwithmu}) is only invariant under transformations that satisfy ${\bf B}M=M{\bf B}$,
which reduces the global symmetry to SU($N_{\! f}$)$_L \times$SU($N_{\! f}$)$_R \times$U(1)$_B$ for $\mu\neq0$. Moreover,
from Eq.\ (\ref{lagwithmu}) it is straightforward to verify that the minimum of the classical potential
is obtained when $\Phi_0 = \phi_0 {\bf \Sigma}$, where $\phi_0^2 = -(m^2-4\mu^2)/(8N_{\! f}\lambda_1+4\lambda_2)$ 
for $c=0$ and $m^2 < 4\mu^2$. 
This expectation value breaks the symmetry down to Sp($N_{\! f}$)$\times$Sp($N_{\! f}$), as required.
It should also be noted that the choice $\Phi_0 = \phi_0 {\bf E}$ now corresponds to a local maximum, so that 
the effective model does distinguish between the two vacua. 

\section{Melting the Diquark Condensate}
\label{III}

As the temperature is increased, at a given $\mu>0$, we expect the
diquark condensate to eventually melt and the symmetry to be
restored. The object is to determine the order of this phase transition
by studying the fixed points of the $\beta$-functions in the $\epsilon$
expansion. 
The effect of the chemical potential in the Landau theory is
threefold: 
{\sl 1)} it allows for terms linear in the time derivative, 
{\sl 2)} it explicitly breaks the U($2N_f$) symmetry, and  
{\sl 3)} it induces a Bose diquark condensate.    

We will assume that $T_c$ at $\mu=0$ is large, $T_c^2\gg |m^2|$.
Then as discussed in detail in \cite{SLW} we can dimensionally reduce
the Landau theory to an effective theory of the massless zeroth
Matsubara modes. It is in this dimensionally reduced theory we will
study the stability of the $\beta$-functions.  Because of the breaking
of the U($2N_f$) symmetry not all zeroth Matsubara modes become
massless at $T_c$ when $\mu>0$. 
At nonzero $\mu$ the effective three dimensional theory describes the
static zeroth Matsubara modes from the block diagonal modes in $\Phi$.
These are the massless modes at the phase transition. 
Explicitly, in the three dimensional theory the field $\Phi$ is 
\be 
\Phi \rightarrow \left ( \matrix{ {\bf A}_1 &
{\bf 0} \cr {\bf 0} & {\bf A}_2 } \right ) \ ,\label{phiformu} 
\ee
with ${\bf A}_1$ and ${\bf A}_2$ $N_{\! f}\times N_{\! f}$ matrices
given by, 
\be 
{\bf A}_1 = \left ( \matrix{ 0 & \phi_{12} & \cdots &
\phi_{1N_{\! f}} \cr -\phi_{12} & 0 & \cdots & \phi_{2N_{\! f}} \cr
\vdots & \vdots & \hbox{} & \vdots \cr -\phi_{1N_{\! f}} &
-\phi_{2N_{\! f}} & \cdots & 0 } \right ) \ \ \ {\rm and} \ \ \ {\bf
A}_2 = \left ( \matrix{ 0 & \phi_{(N_{\! f}+1)(N_{\! f}+2)} & \cdots &
\phi_{(N_{\! f}+1)(2N_{\! f})} \cr -\phi_{(N_{\! f}+1)(N_{\! f}+2)} &
0 & \cdots & \phi_{(N_{\! f}+2)(2N_{\! f})} \cr \vdots & \vdots &
\hbox{} & \vdots \cr -\phi_{(N_{\! f}+1)(2N_{\! f})} & -\phi_{(N_{\!
f}+2)(2N_{\! f})} & \cdots & 0}\right ) \,\, .  
\ee 
When the chemical potential is infinitesimally small, the masses of
the zeroth Matsubara components of the off-diagonal blocks in $\Phi$
are not large. In this particular situation, the phase transition 
is almost identical to the phase transition in the theory for $\mu=0$,
at least if it is first 
order \cite{wirstam}. We therefore restrict $\mu$ to be sufficiently
large for this separation of mass scales to hold, although it should
be emphasized that the matching to the nonzero $T$, $\mu =0$, line in
the phase diagram may be somewhat ambiguous.

Our approach is to use renormalization group arguments in 
$(4-\epsilon)$ dimensions and study the existence of infra-red fixed
points in the space of coupling constants for $\epsilon
\rightarrow 1$. Whether or not this extrapolation is valid is of
course a difficult question to answer fully. However, this approach
has been successful under many different circumstances, both in the
condensed matter context (see e.g. \cite{justin,GZ-J} and references
therein) and in the context of relativistic field theories
\cite{robfrank,electroweak}.

\subsection{Renormalization group flow in the anomaly-free case}

Having defined the model, we now analyze the renormalization group flow
in the space of parameters $\lambda_1$, $\lambda_2$, and $c$.  
As long as the quartic coupling constants are within their stability
region, two situations can occur. Either the $\beta$-functions have
infrared stable fixed points, in which case there is a second order
phase transition, or they do not, and the phase transition is then
first order, induced by fluctuations. The obvious caveat here is that
the coupling constants can be adjusted to lie outside the stability
region, with the result that higher dimensional terms must be added to
stabilize the theory, making the phase transition first order
already at the mean field level. In addition, there could also be
fixed points at the nonperturbative level.

From the above discussion, the $c=0$ effective Lagrangian 
around criticality is
\be
L = \frac{1}{2} {\rm Tr}\left [ (\del_{\nu}\Phi^{\dagger} ) (\del^{\nu}\Phi ) \right ]
-\lambda_1 \bigl ( {\rm Tr} \left [ \Phi^{\dagger} \Phi \right ]
\bigr )^2 - \lambda_2 {\rm Tr} \bigl [ \left (\Phi^{\dagger} \Phi \right )^2
\bigr ] \ ,
\label{lagrattc}
\ee
where $\Phi$ is defined as in Eq.\ (\ref{phiformu}). 
To study the existence of fixed points, we expand the Lagrangian
in Eq.\ (\ref{lagrattc}) to one-loop order around the expectation value
$\Phi_0 =\phi_0{\bf \Sigma}$ and calculate the $\beta$-functions using
the background field method. To more easily separate
the one-loop corrections to $\lambda_1$ and $\lambda_2$, we will 
also label
the constant $\phi_0$ differently for different matrix elements:
\be
\Phi_0 \rightarrow  i \left ( \matrix{ \phi_a{\bf I} & {\bf 0} \cr
{\bf 0} & \phi_b{\bf I} } \right ) \ .
\ee
We then replace  $\Phi\rightarrow \Phi_0 + \Phi$ in Eq.\ (\ref{lagrattc}) 
and split the fields into their real and imaginary parts, 
$\phi_{mn}=(\phi_{mn}^{(r)}+i \phi_{mn}^{(i)})/\sqrt{2}$. 
Up to terms quadratic in the fluctuations, we find that
\be
&&L = -V_0 +\frac{1}{2}\sum_{k,l=1}^{N_{\! f}}\biggl . \biggl [ (\del_{\nu}\phi_{kl}^{(r)} )^2 
+(\del_{\nu}\phi_{kl}^{(i)} )^2 + (\del_{\nu}\phi_{(k+N_{\! f})(l+N_{\! f})}^{(r)} )^2 +
(\del_{\nu}\phi_{(k+N_{\! f})(l+N_{\! f})}^{(i)} )^2 \biggr ] \biggr |_{k<l} \nonumber \\ &&-
\lambda_1 \biggl [ 8\phi_a^2 \biggl (\sum_{k=1}^{N_{\! f}/2}\phi_{k(N_{\! f}/2+k)}^{(i)} \biggr 
)^2+8\phi_b^2 \biggl (\sum_{k=N_{\! f}+1}^{3N_{\! f}/2}\phi_{k(N_{\! f}/2+k)}^{(i)} \biggr
)^2 +16\phi_a\phi_b \biggl (\sum_{k=1}^{N_{\! f}/2}\phi_{k(N_{\! f}/2+k)}^{(i)} \biggr   
)\!\! \biggl (\sum_{k=N_{\! f}+1}^{3N_{\! f}/2}\phi_{k(N_{\! f}/2+k)}^{(i)} \biggr ) \biggr . \nonumber \\ 
&&+ \biggl . 2N_{\! f}(\phi_a^2+\phi_b^2) \biggl (\sum_{k,l=1}^{N_{\! f}} \biggl .
\biggl \{ (\phi_{kl}^{(r)})^2 +(\phi_{kl}^{(i)})^2 +(\phi_{(N_{\! f}+k)(N_{\! f}+l)}^{(r)})^2
+(\phi_{(N_{\! f}+k)(N_{\! f}+l)}^{(i)})^2 \biggr \} \biggr |_{k<l}\biggr ) \biggr ] \nonumber \\
&&-\lambda_2 \biggl [4\phi_a^2\sum_{k,l=1}^{N_{\! f}} \! \! \biggl .
\{ (\phi_{kl}^{(r)})^2 +(\phi_{kl}^{(i)})^2 \}\biggr |_{k<l} \! + 4\phi_b^2\!\!\!\!\sum_{k,l=N_{\! f}+1}^{2N_{\! f}} \!\!\!
\biggl . \{ (\phi_{kl}^{(r)})^2 +(\phi_{kl}^{(i)})^2 \} \biggr |_{k<l} \!-2\phi_a^2 \biggl \{
\sum_{k=1}^{N_{\! f}/2} \biggl \{ (\phi_{k(N_{\! f}/2+k)}^{(r)})^2 -(\phi_{k(N_{\! f}/2+k)}^{(i)})^2 \biggr \} \biggr .
\biggr . \nonumber \\ &&+ \biggl . 2\sum_{k,l=1}^{N_{\! f}/2} \biggl . \biggl
(\phi_{k(N_{\! f}/2+l)}^{(r)}\phi_{l(N_{\! f}/2+k)}^{(r)}-\phi_{k(N_{\! f}/2+l)}^{(i)}\phi_{l(N_{\! f}/2+k)}^{(i)}+\phi_{kl}^{(r)}\phi_{(N_{\! f}/2+k)(N_{\! f}/2+l)}^{(r)}
-\phi_{kl}^{(i)}\phi_{(N_{\! f}/2+k)(N_{\! f}/2+l)}^{(i)} \biggr ) \biggr |_{k<l}\biggr \}\nonumber \\
&&-2\phi_b^2 \biggl \{ 2\sum_{k,l=N_{\! f}+1}^{3N_{\! f}/2} \biggl . \biggl
(\phi_{k(N_{\! f}/2+l)}^{(r)}\phi_{l(N_{\! f}/2+k)}^{(r)}-\phi_{k(N_{\! f}/2+l)}^{(i)}\phi_{l(N_{\! f}/2+k)}^{(i)}+
\phi_{kl}^{(r)}\phi_{(N_{\! f}/2+k)(N_{\! f}/2+l)}^{(r)}
-\phi_{kl}^{(i)}\phi_{(N_{\! f}/2+k)(N_{\! f}/2+l)}^{(i)} \biggr ) \biggr |_{k<l}\biggr . \nonumber \\ &&+ 
\biggl . \biggl . 
\sum_{k=N_{\! f}+1}^{3N_{\! f}/2} \biggl \{ (\phi_{k(N_{\! f}/2+k)}^{(r)})^2 -(\phi_{k(N_{\! f}/2+k)}^{(i)})^2 \biggr \}
\biggr \} \biggr ] \ ,\label{quadraticl}
\ee
where the classical potential $V_0$ is given by,
\be
V_0 = \lambda_1N_{\! f}^2(\phi_a^2 +\phi_b^2)^2+\lambda_2N_{\! f}(\phi_a^4+\phi_b^4) \ . \label{classicalpot}
\ee

All the terms quadratic in $\phi_a^2$ and $\phi_b^2$ can be collected
into a symmetric matrix $M$ that can be diagonalized by an orthonormal
transformation, $M\rightarrow PMP^T=M_D$. By integrating out the
quadratic fluctuations and using dimensional regularization in
$d=4-\epsilon$ dimensions, we find the effective action,
\be
V_{{\rm eff.}}(\phi_a,\phi_b) = V_0 -\frac{1}{32\pi^2\epsilon}\left ({\rm
Tr}M_D^2 \right ) +\delta\lambda_1N_{\! f}^2(\phi_a^2 +\phi_b^2)^2+\delta
\lambda_2N_{\! f}(\phi_a^4+\phi_b^4)\ , \label{effectivepot} 
\ee
where we have only kept the ultra-violet divergent terms
and $\delta\lambda_1$  and $\delta\lambda_2$ denote the counterterms.
Using the relation ${\rm Tr}M_D^2 = {\rm Tr}M^2$, it is
straightforward to find the counterterms and hence 
the $\beta$-functions.  Using
$\kappa$ as the arbitrary mass scale, we find 
\be 
\beta_1 = &&\kappa\frac{\del\lambda_1}{\del\kappa} = -\epsilon\lambda_1 +\frac{1}{\pi^2}\biggl [ 
\lambda_1^2(N_{\! f}^2-N_{\! f}+4) +2\lambda_1\lambda_2(N_{\! f}-1)\biggr ]\label{betafunctions1} \ , \\
\beta_2 = &&\kappa\frac{\del\lambda_2}{\del\kappa} = -\epsilon\lambda_2 +\frac{1}{2\pi^2}\biggl [
5\lambda_2^2(N_{\! f}-1)+12\lambda_1\lambda_2 \biggr ] \ . \label{betafunctions2}
\ee

Consider now the different fixed points that can be inferred from 
Eqs.\ (\ref{betafunctions1})--(\ref{betafunctions2}), i.e. 
the solutions to $\beta_1=\beta_2=0$. Since we are examining the
long-range fluctuations, the fixed points $\lambda_1^{\ast}$ 
and $\lambda_2^{\ast}$ must be infra-red stable, 
implying that the eigenvalues of the stability matrix,
\be
S= \biggl . \biggl ( \matrix{ \del\beta_1/\del\lambda_1 & \del\beta_1/\del\lambda_2 
\cr \del\beta_2/\del\lambda_1 & \del\beta_2/\del\lambda_2 } \biggr ) 
\biggr |_{\stackrel{{\scriptstyle \lambda_1=\lambda_1^{\ast}}}{\lambda_2=\lambda_2^{\ast}}} \ , \label{stabmatrix}
\ee
have to be positive. From 
Eqs.\ (\ref{betafunctions1})--(\ref{betafunctions2}), 
we find the fixed points 
\be
&&{\rm i:}\ \ \lambda_1^{\ast}=0 \ , \ \ \lambda_2^{\ast}=0 \ , \hspace{3.05cm}
{\rm ii:}\ \ \lambda_1^{\ast}=0 \ , \ \ \lambda_2^{\ast}=\frac{2\pi^2\epsilon}{5(N_{\! f}-1)} \ , 
\nonumber \\ &&{\rm iii:}\ \ \lambda_1^{\ast}=\frac{\pi^2\epsilon}{N_{\! f}^2-N_{\! f}+4} \ , \ \ \lambda_2^{\ast}=0 \ , 
\hspace{1cm} {\rm iv:}\ \ \lambda_1^{\ast}=\frac{\pi^2\epsilon}{5N_{\! f}(N_{\! f}-1)-4} \ , \ \ \lambda_2^{\ast}=
\frac{2\pi^2\epsilon(N_{\! f}^2-N_{\! f}-2)}{(N_{\! f}-1)[5N_{\! f}(N_{\! f}-1)-4]} \ ,    
\ee
where the last one (iv) only exists for $N_{\! f}\geq 4$. In fact, 
when $N_{\! f}=2$ it reduces to the third solution.
From the eigenvalues of $S$, we find that the first and second 
cases are unstable for any number
of flavors. The third case is unstable for $N_{\! f}\geq 4$ 
but marginal when $N_{\! f}=2$, i.e. one of the 
eigenvalues vanishes. The fourth solution, however, has a 
fixed point that is infra-red stable for $N_{\! f}\geq 4$.
From this we conclude
\begin{itemize}

\item For $N_{\! f}=2$, the order of the phase transition is 
not determined by the renormalization group flow, and
a higher order calculation is 
necessary to make a definite prediction. However, at $\mu =0$, the 
temperature induced phase transition
is first order for $N_{\! f}=2$ and $c=0$ \cite{wirstam},
so it is likely 
that this first order transition extends to small but nonzero
values of $\mu$. If the phase transition becomes second order as $\mu$ 
increases further, it should correspond to the O(4) critical point (i.e. case
iii above). 

\item For $N_{\! f}\geq4$, there is a stable fixed point. Thus, if
this theory evolves  
from a point within the stability region  $\lambda_2 \geq 0$ and
$\lambda_1 +\lambda_2/(2N_{\! f})\geq 0$, 
the phase transition in two--color QCD with more than four flavors,
and without any influence from the axial anomaly, is likely to be second order.

\end{itemize}

\begin{figure}
\epsfxsize=3in
\centerline{\epsfbox{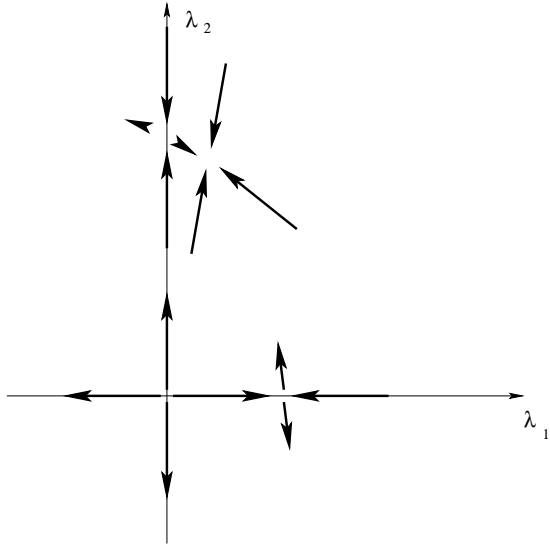}}
\caption[]{The flow in the $(\lambda_1,\lambda_2)$-plane for $N_f=4$
and without the anomaly term. 
The classical potential is stable for $\lambda_2\geq0$ and
$\lambda_1+\lambda_2/(2N_f)\geq0$.
The vectors shown have a length given by
the magnitude of the eigenvalues of the stability matrix at the fixed
point. The direction is away from the fixed point for negative
eigenvalues and towards the fixed point for positive eigenvalues. The
orientation of the vectors is given by the corresponding eigenvectors
of the stability matrix. There is a stable fixed point and the phase
transition is predicted to be of second order. For $N_f>4$ the flow
diagram has a similar topology. For $N_f=2$ the upper
right hand stable fixed point merges with the unstable fixed point on
the $\lambda_1$ axis and becomes marginally stable.}
\label{fig:flow}
\end{figure}

The renormalization group flow in the $(\lambda_1,\lambda_2)$-plane for
$N_f=4$ is illustrated in figure \ref{fig:flow}.

\subsection{Adding the chiral anomaly}

The above conclusions must be reconsidered if the chiral anomaly
is present around $T_c$.  Lattice results for $N_c=3$ QCD
indicate that although the effects of the chiral anomaly decrease,
they do not vanish in the vicinity of the chiral symmetry 
restoring phase transition \cite{anomalylattice}. 
Since there is no reason to believe that the
situation is drastically different in $N_c=2$ QCD, the anomaly can
be expected to be present near $T_c$ and $\mu = 0$. At a nonzero
chemical potential, there is an additional suppression of the anomaly
\cite{anomalysup}  apart from the nonzero $T$ effects. However, since
our model is limited to the region $\mu\lton \Lambda_{{\rm
QCD}}$, it is still likely that the effects of the chiral anomaly 
are not negligible at the phase transition. 

For two flavors, the anomaly term is a mass-like operator  
and has the effect that only half of the number of the 
original real-valued fields in
${\bf A}_1$ and ${\bf A}_2$ become critical. At $T=0$, there is an
SU(2)$\times$SU(2)$\times$U(1) symmetry that is spontaneously broken
to Sp(2)$\times$Sp(2), but since Sp(2)$\sim$SU(2), only the
U(1)$\sim$O(2) symmetry is actually broken.  
Keeping only those fields in Eq.\ (\ref{lagwithmu}) which become
critical at $T_c$, the theory becomes O(2)-symmetric and the diquark
condensate breaks this symmetry completely.  Thus, if the phase transition 
indeed is second order, it is characterized by O(2) critical exponents. Since the
phase transition at $\mu =0$ can be second order as well
\cite{wirstam}, but with O(6) critical exponents, there could thus be
a change of the critical exponents as $\mu$ increases from zero.

For $N_{\! f}=4$, the anomaly term represents another
quartic coupling.  Repeating the calculation from the previous section
with $c\neq 0$, we find the $\beta$-functions,
\be 
\beta_1 = &&\kappa\frac{\del\lambda_1}{\del\kappa} = -\epsilon\lambda_1
+\frac{1}{\pi^2}\biggl [ 16\lambda_1^2 +6\lambda_1\lambda_2
+\frac{c^2}{64} \biggr ] \ , \nonumber \\ 
\beta_2 = &&\kappa\frac{\del\lambda_2}{\del\kappa} = -\epsilon\lambda_2
+\frac{1}{\pi^2}\biggl [ \frac{15\lambda_2^2}{2} +6\lambda_1\lambda_2
+\frac{c^2}{32} \biggr ] \ , \nonumber \\ \beta_3 = &&\kappa\frac{\del
c}{\del\kappa} = -\epsilon c +\frac{2c}{\pi^2}\biggl [ 3\lambda_1
-\lambda_2 \biggr ] \ . \label{betafunctionswithc} 
\ee 
When the fixed point $c^{\ast}$ vanishes, the solutions of course 
reduce to the results of the previous section.
Surprisingly, there are no real-valued solutions to the three equations that satisfy
$c^{\ast}\neq 0$.  Since there is then no fixed point, we conclude
that the phase transition is first order, induced by fluctuations
\cite{amit}.  The presence of a first order phase transition for
$N_{\! f}=4$ is also in agreement with the existing lattice results
\cite{Kogut:2001if}.
Note, however, (see Ref.\ \cite{peskin,Hands:2000ei}) that the roles of the fundamental and
the adjoint representations are reversed for the staggered fermions as used in
Ref.\ \cite{Kogut:2001if}. Hence, this comparison is not fully justified.

For $N_{\! f}\ge6$, the critical behavior in the presence of
the chiral anomaly is identical to the $c=0$ case. In renormalization
group terms, the operator $c$ is irrelevant, since its mass dimension
in four Minkowski space-time dimensions is $[c]= 4-N_{\!  f}$, and the
critical behavior is determined by $\lambda_1$ and $\lambda_2$ only.

Finally, we stress again that in this approach we can not
exclude first order phase transitions induced dynamically by higher order
operators. Hence, even if there is a stable fixed point within the
$\epsilon$-expansion, the transition may still be first order.

\section{Conclusions}
\label{V}

In this work, we have studied the diquark phase transition in two-color
QCD at
zero quark mass and nonzero $T$ and $\mu$ using an effective Landau
theory. To improve upon mean-field theory, we studied the renormalization group
flow in $4-\epsilon$ dimensions, and then extrapolated our results to
$\epsilon =1$.  
The conclusions to be drawn are the following. In the absence of the
anomaly term and for $N_{\! f}\geq 4$, the phase transition could be 
second order, whereas the case of $N_{\! f}=2$ is inconclusive at the one-loop
level studied here.
The effect of the anomaly on the phase transition has been shown to be
substantial. In the presence of the anomaly, the $N_{\! f}=2$ case exhibits a 
second order transition characterized by O(2) critical exponents,
which may change to O(6) as $\mu \rightarrow 0$.  
For $N_{\! f}=4$, there is no infrared stable fixed points and 
hence the phase transition is first order.  For $N_{\! f}\geq6$, the phase 
transition may be of second order, since this situation is identical
to the $c=0$ case. 
Of course, the statement for $N_{\! f}\geq 6$
is only true as long as $N_c=2$ QCD is asymptotically free. From the one-loop
running coupling  constant, this limits the number of flavors to $N_{\! f} \leq
N_{\! f\, ({\rm max})} = 11$.
 
Our analysis has focused solely on the case of massless quarks, and it
is interesting to speculate on the consequences of adding nonzero
masses. Defining $x=2\mu /m_{\pi} \propto \mu /
\sqrt{m_q}$, the phase transition discussed in this manuscript
corresponds to the limit $x \rightarrow \infty$. 
For a common, nonzero quark mass $m_q$, the diquark condensate
competes with the chiral condensate. 
Such a term alters the
symmetry breaking pattern from 
SU($N_{\! f}$)$_L\times$SU($N_{\! f}$)$_R\times$U(1)$_B\rightarrow$Sp($N_{\!
  f}$)$\times$Sp($N_{\! f}$)  
in the massless case to SU($N_{\! f}$)$_V\times$U(1)$_B\rightarrow$Sp($N_{\! f}$) 
when $m_q\neq 0$.  For $\mu\gg m_\pi$, the first order phase transition for 
$N_{\! f} =4$ and $c\neq0$ should only weaken, whereas the second order 
phase transitions for $N_{\! f} \geq 6$ presumably stay second order. 
When $ N_{\! f} =2$ the number of broken
symmetries is independent of the introduction of a quark mass, since only the
baryon number symmetry is broken. It is therefore possible that the phase
transition is second order for all $x$. For $N_{\! f}\geq 4$, however, it is
unlikely that the massless results should be 
representative at such a small ratio of $\mu$ and $m_{\pi}$. Instead, the
mass term should be incorporated explicitly into the effective Lagrangian, 
and the effects could very well change the order of the phase
transition as $x$ approaches $1$.

The presence of a tricritical point in the phase diagram of two color
QCD has so far only been observed on the lattice for $N_{\! f}=4$ flavors. Chiral 
perturbation theory \cite{STV2} predicts that the tricritical point is present 
for any even number of flavors ($N_{\! f} \leq N_{\! f\, ({\rm max})} = 11$). 
In contrast, the possibility of a fluctuation induced first order phase
transition depends very sensitively on the number of flavors as well as 
the strength of the anomaly.  By combining the results of future lattice 
simulations for different numbers of flavors with these complementary 
but quite different analytical approaches, the nature of the superfluid 
phase transition will certainly become more clear.

In recent lattice investigations \cite{Kogut:2002tm,KS2}, 
it was found that the phase
diagram of $N_c=3$ QCD at nonzero temperature and isospin chemical 
potential is very similar to that of $N_c=2$ at nonzero temperature 
and baryonic chemical potential. In particular, the existence of 
a tricritical point was also demonstrated.
Chiral perturbation theory also predicts the existence of a tricritical 
point for $N_c=3$ QCD at nonzero isospin chemical potential \cite{STV2}. 
The approach followed in this work has been applied to this 
physically relevant case \cite{SLW} and for $N_f=2$ one marginal 
fixed point was found for the anomaly free case.  

By carefully combining the analytical and numerical results for a
variety of parameters (including not the least of which is the case where
$N_{\! f}$ is odd), we should ultimately be able to grasp the complex phase
diagram of $N_c=2$ QCD. Hopefully, this experience will also further 
enhance our knowledge of dense $N_c=3$ QCD.

\begin{center}
{\bf Acknowledgments}
\end{center}
\noindent
The authors would like to thank R.\ D.\ Pisarski for valuable and
encouraging discussions and for reading the manuscript. 
F.\ Sannino is thanked for collaboration during the early stages of this work.
We would also like to thank A.\ Krasnitz and M.\ Oswald for useful discussions
and the Institut d'{\'E}tudes Scientifiques de Carg{\`e}se where this
work was initiated. 
J.\ W.\ would like to thank Brookhaven National Laboratory, where part
of this work was done with
support from The Swedish Foundation for International
Cooperation in Research and Higher Education (STINT) under contract 99/665,
and the DOE grant DE-AC02-98CH10886. J.T.L.\ is supported by 
the U.S.\ DOE grant DE-FG02-97ER41027.
K.S.\ was supported by the Leon Rosenfeld Foundation and by the 
U.S.\ DOE grant DE-FG02-88ER40388.

\end{document}